\documentclass{PoS}

\title{The nucleon's transversity and the photon's distribution amplitude probed in lepton pair photoproduction}

\ShortTitle{The nucleon's transversity and the photon's DA}

\author{B. Pire\\
        CPhT, \'Ecole Polytechnique, CNRS, 91128 Palaiseau, France\\
}

\author{\speaker{L. Szymanowski}
\\
        SINS, Warsaw, Poland\\
        E-mail: \email{lechszym@fuw.edu.pl}}

\abstract{We describe a new way to access the chiral odd transversity parton distribution in the proton through
the photoproduction of lepton pairs on transversely polarized target. The basic ingredient is the interference of the Bethe
Heitler or Drell-Yan amplitudes with the amplitude of a process, where the photon couples to quarks
through its twist-2 chiral-odd distribution amplitude. This approach permits to scan experimentally the  shape of the transversity distribution in a nucleon as well as the photon distribution amplitude.  }

\FullConference{35th International Conference of High Energy Physics - ICHEP2010,\\
		July 22-28, 2010\\
		Paris France}

\begin{document}

Transversity quark distributions in the nucleon remain among the most unknown leading twist hadronic observables. This is mostly due to their chiral odd character which enforces their decoupling in most hard amplitudes. After the pioneering studies \cite{tra}, much work \cite{Barone} has been devoted to the exploration of many channels but experimental difficulties have challenged the most promising ones. The recent focuses on transversity dependent observables in single inclusive deep inelastic scattering propose to assume the factorization of transverse momentum dependent parton distributions (TMDs) and fragmentation functions. This allowed  some first extraction of chiral odd quantities,  demonstrating, although  in a weak sense, their non-zero value. Here we restrict on the use of leading twist factorizable quantities, such as integrated parton distributions and distribution amplitudes.

A new way to access the chiral odd transversity parton distribution in the proton emerges thanks to the observation that the photon twist 2 distribution amplitude (DA) \cite{Braun}  is chiral-odd.  This latter object is normalized to the magnetic susceptibility of the QCD vacuum. It is defined as
\begin{eqnarray}\label{def3:phi}
\langle 0 |\bar q(0) \sigma_{\alpha\beta} q(x) 
   | \gamma^{(\lambda)}(k)\rangle 
=       
 i \,e_q\, \chi\, \langle \bar q q \rangle
 \left( \epsilon^{(\lambda)}_\alpha k_\beta-  \epsilon^{(\lambda)}_\beta k_\alpha\right)  
 \int\limits_0^1 \!dz\, e^{-iz(kx)}\, \phi_\gamma(z)\,,
\label{phigamma}
\end{eqnarray}    
where the normalization is chosen as $\int dz\,\phi_\gamma(z) =1$, 
and $z$ stands for the momentum fraction carried by the quark. The product of the quark condensate and of the magnetic susceptibility of the QCD vacuum
$\chi\, \langle \bar q q \rangle$ has been estimated  to be of the order of 50 MeV.  

\begin{figure}[h]
\begin{center}
\includegraphics[width=4.8cm]{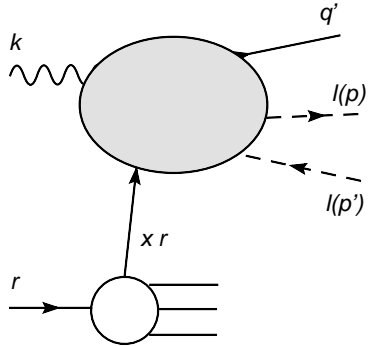}
\caption{The partonic version of the process $\gamma(k,\epsilon)N(r,s_T) \to l^-(p)l^+(p')X(=q')$}
\label{fig1}
\end{center}
\end{figure}

\begin{figure}[h]
\begin{center}
\includegraphics[width=3.5cm]{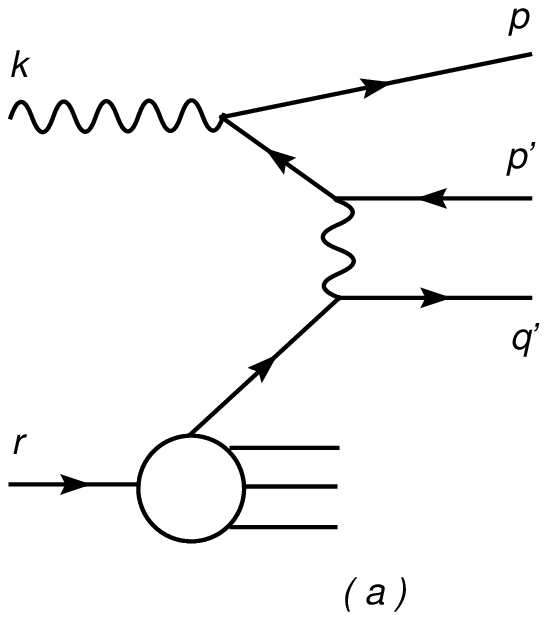}
\includegraphics[width=3.5cm]{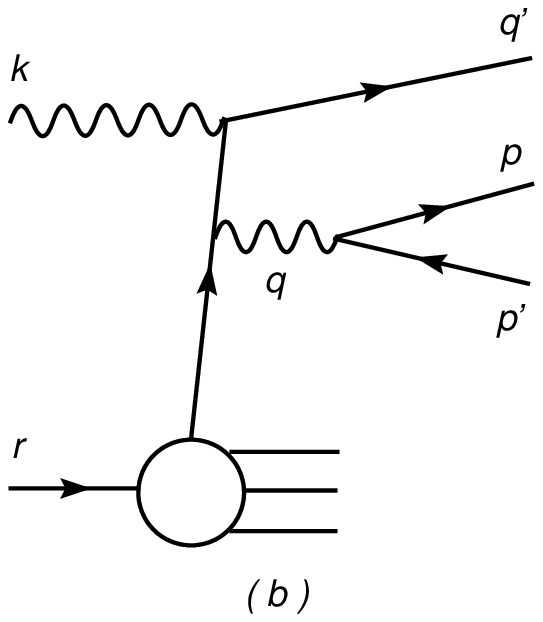}
\includegraphics[width=3.5cm]{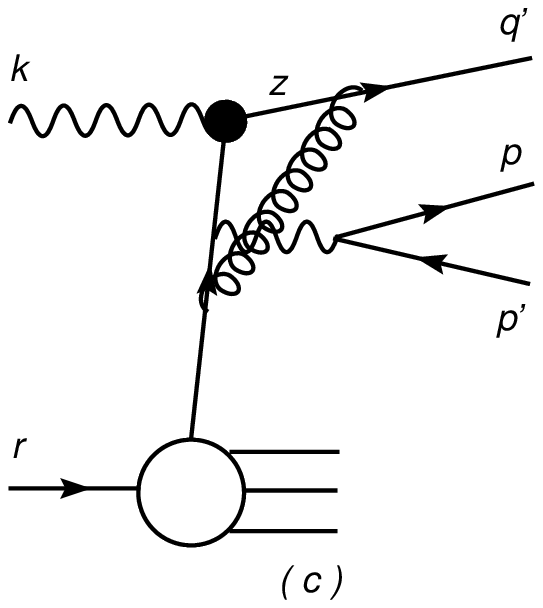}
\includegraphics[width=3.5cm]{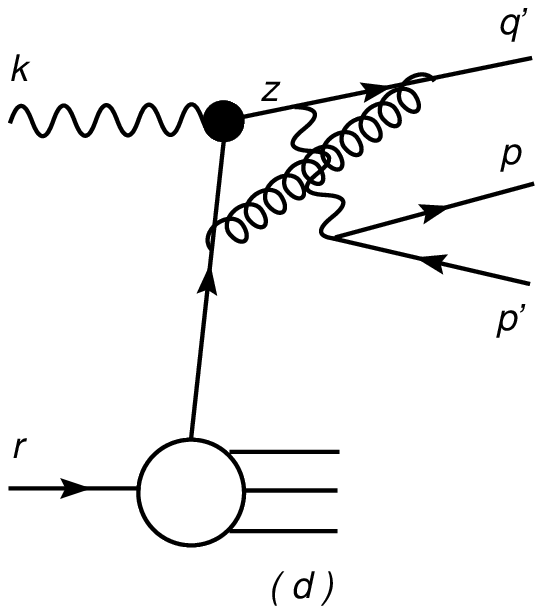}
\caption{Some amplitudes contributing to lepton pair photoproduction. (a) : The Bethe-Heitler process. (b) : The Drell-Yan process with the photon pointlike coupling. (c) -(d) : The Drell-Yan process with the photon Distribution Amplitude. }
\label{fig2}
\end{center}
\end{figure}

The basic ingredient of our proposal \cite{PS} is the interference of the Bethe Heitler or Drell-Yan amplitudes with the amplitude of a process, where the photon couples to quarks through this chiral-odd DA. We thus  consider the  process illustrated in Fig.~\ref{fig1}:
\begin{equation}
\label{process}
\gamma(k,\epsilon) N (r,s_T)\to l^-(p)  l^+(p') X\,,
\end{equation}
with $q= p+p'$ in the kinematical region where $Q^2=q^2$ is large and the transverse component $ |\vec Q_\perp |$ 
of $q$ is of the same order as $Q$. $s_T$ is the transverse polarization vector of the nucleon. 
Fig.~\ref{fig1} shows a partonic version of the process (\ref{process}) in which the striked quark carries  
fraction $x$ of nucleon's momentum and which leads to the final state containing a lepton pair and a jet $q'$ which balances the lepton pair momentum.
Such a process  occurs either through a Bethe-Heitler amplitude (Fig.~\ref{fig2}a) where the initial photon 
couples to a final lepton, or through Drell-Yan type amplitudes (Fig.~\ref{fig2}b) where the final leptons originate from 
a virtual photon. Among these Drell-Yan processes, one must distinguish the cases where the real photon couples 
directly (through the QED coupling) to quarks or through its quark content, {\em i.e.} the photon structure function. 
Gluon radiation at any order in the strong coupling $\alpha_s$, does not  introduce any chiral-odd 
quantity if one neglects quark masses. We next consider the  contributions where the photon couples to the strong 
interacting particles through its twist-2 distribution amplitude (Figs.~\ref{fig2}c and \ref{fig2}d). 
One can easily see by inspection that this  is the only way to get at the level of twist 2 (and with vanishing quark masses) 
a contribution to  nucleon transversity dependent observables.
We call this amplitude ${\cal A}_\phi$ :
\begin{eqnarray}\label{AmpC}
{\cal A}_\phi (\gamma q \to l \bar l q) = 2i \frac{C_F}{4N_c} e_q^2 e 4\pi\alpha_s \frac{ \chi\, \langle \bar q q \rangle }
{Q^2} \int dz \phi_\gamma (z)  \bar u(q') [ \frac{A_1}{x\bar z s (t_1+i\epsilon)} +\frac{A_2}{ z u (t_2+i\epsilon)} ]u(r) \bar u(p)\gamma^\mu v(p') \,,\nonumber
\end{eqnarray} 
with  $t_1= (zk-q)^2$ and $t_2= (\bar z k -q)^2$.
$
A_1 = x \,\hat r\,\hat \epsilon\,\hat k \,\gamma^\mu + \gamma^\mu \,\hat k \,\hat \epsilon \,\hat q $ and
$A_2 = \hat \epsilon\,\hat q\, \gamma^\mu \,\hat k + \hat k \,\gamma^\mu \,\hat q \,\hat \epsilon$
 do not depend on the light-cone fraction $z$.
${\cal A}_\phi$ develops an absorptive part  proportional to
\begin{eqnarray}
\label{abs}
\int dz \phi_\gamma (z) \bar u(q') [\frac{A_1}{x\bar z s}\delta (t_1) +\frac{A_2}{ z u} \delta (t_2)]u(r) \bar u(p)\gamma^\mu v(p')\,. \nonumber
\end{eqnarray}
This allows to perform the $z-$integration, the result of which, after using the $z-\bar z$ symmetry of the distribution amplitude, yields an absorptive part of the  amplitude ${\cal A}_\phi$ proportional to 
$\phi_\gamma (\frac{\alpha Q^2}{ Q^2+\vec Q_\perp ^2})$, where $\alpha$ is the component along the photon momentum $k$ of the lepton pair momentum $q$, see \cite{PS}. This  absorptive part, which may be measured in single spin asymmetries, as discussed below, thus scans the photon chiral-odd distribution amplitude.

\begin{figure}[h]
\begin{center}
\includegraphics[width=7.4cm]{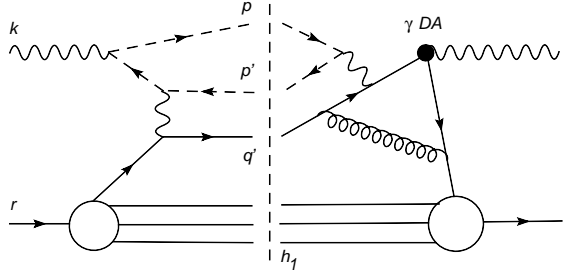}
\caption{Interference of the BH contribution with the amplitude involving the chiral-odd photon DA}
\label{fig3}
\end{center}
\end{figure}

The cross section for reaction (\ref{process}) can  be read as
\begin{eqnarray}\label{cs}
\frac {d\sigma}{d^4Q \,d\Omega}  =  \frac {d\sigma_{BH}}{d^4Q\,d\Omega} +  \frac {d\sigma_{DY}}{d^4Q\,d\Omega} +   \frac {d\sigma_{\phi}}{d^4Q\,d\Omega} +  \frac {\Sigma d\sigma_{int}}{d^4Q\,d\Omega}\,,\nonumber
\end{eqnarray}
where $\Sigma d\sigma_{int}$ contains various interferences, while
the transversity dependent  differential cross section (we denote $\Delta_T \sigma = \sigma(s_T) - \sigma(-s_T)$) reads
\begin{eqnarray}\label{cst}
&&\frac {d\Delta_T \sigma}{d^4Q\,d\Omega}  = \frac {d\sigma_{\phi int}}{d^4Q\,d\Omega}  \,,
\end{eqnarray}
 where $d\sigma_{\phi int}$ contains only interferences between the amplitude ${\cal A}_\phi$ and the other amplitudes. Moreover, one may use the distinct charge conjugation properties (with respect to the produced lepton part) of the Bethe Heitler
 amplitude (Fig.~\ref{fig2}a) versus the Drell-Yan amplitude (Fig.~\ref{fig2}b). The charge asymmetry
  selects the interference between ${\cal A}_\phi$ and the Bethe-Heitler amplitude :
\begin{eqnarray}\label{CAcst}
&&\frac {d\Delta_T \sigma (l^-) - d\Delta_T \sigma (l^+) }{d^4Q\,d\Omega}  = \frac {d\sigma_{\phi BH}}{d^4Q\,d\Omega}  \,.
\end{eqnarray}
as illustrated in Fig.~\ref{fig3} (plus its complex conjugate).
The polarization average of $d\sigma_{\phi BH}$, i.e.
$\frac {1}{2} \sum_\lambda d\sigma_{\phi BH} (\gamma(\lambda)p\to l^-l^+X)
$
is proportional to the product
$
\phi_\gamma [\frac{\alpha Q^2}{ Q^2+\vec Q_\perp ^2}]\cdot h_1^q(\frac{Q^2}{\alpha s}+\frac{\vec Q_\perp ^2}{\alpha \bar\alpha s} )
$
of the photon distribution amplitude $\phi_\gamma$ and the transversity distribution $h_1^q$. This product depends only on the external kinematical variables of the process (\ref{process}). We are thus in a position to scan either the transversity quark distribution or the photon distribution amplitude. 

\vskip.1in
\noindent
 {\bf Acknowledgments}

\noindent 
This work is partly supported by the Polish Grant N202 249235.

\end{document}